\documentclass[aps, prd, floatfix, twocolumn, superscriptaddress, nofootinbib]{revtex4-1}

\usepackage{latexsym}
\usepackage{amsmath}
\usepackage{amssymb}
\usepackage{amsfonts}

\usepackage{slashed}
\usepackage{bm}

\usepackage[vcentermath]{youngtab}

\usepackage{color}
\definecolor{purple}{rgb}{0.5,0,0.5}
\definecolor{blue}{rgb}{0.0,0,0.9}
\definecolor{prdblue}{rgb}{0.133,0.118,0.498}

\usepackage{enumitem}
\usepackage{multirow}
\usepackage{subfigure}
\usepackage{textcomp}

\usepackage{supertabular} 
\usepackage{placeins}
\usepackage{epsfig}
\usepackage{graphicx}

\usepackage[hyperindex,breaklinks]{hyperref}

\hypersetup{
  colorlinks=true,
  linkcolor=blue,
  filecolor=magenta,
  urlcolor=blue,
  citecolor=blue,
}

\begin{document}


\title{Nature of the doubly-charmed tetraquark $T_{cc}^+$ in a constituent quark model}

\author{P. G. Ortega}
\email[]{pgortega@usal.es}
\affiliation{Departamento de Física Fundamental, Universidad de Salamanca, E-37008 Salamanca, Spain}
\affiliation{Instituto Universitario de F\'isica 
Fundamental y Matem\'aticas (IUFFyM), Universidad de Salamanca, E-37008 Salamanca, Spain}

\author{J. Segovia}
\email[]{jsegovia@upo.es}
\affiliation{Departamento de Sistemas Físicos, Químicos y Naturales, Universidad Pablo de Olavide, E-41013 Sevilla, Spain}

\author{D. R. Entem}
\email[]{entem@usal.es}
\affiliation{Departamento de Física Fundamental, Universidad de Salamanca, E-37008 Salamanca, Spain}
\affiliation{Instituto Universitario de F\'isica
Fundamental y Matem\'aticas (IUFFyM), Universidad de Salamanca, E-37008 Salamanca, Spain}

\author{F. Fern\'andez}
\email[]{fdz@usal.es}
\affiliation{Instituto Universitario de F\'isica 
Fundamental y Matem\'aticas (IUFFyM), Universidad de Salamanca, E-37008 Salamanca, Spain}

\date{\today}

\begin{abstract}

The recently discovered $T_{cc}^+$ is evaluated as a $DD^*$ molecular structure in the $J^P=1^+$ sector. A coupled-channels calculation in charged basis, considering the $D^0D^{*\,+}$, $D^+D^{*\,0}$ and $D^{*\,0}D^{*\,+}$ channels, is done in the framework of a constituent quark model that successfully described other molecular candidates in the charmonium spectrum such as the $X(3872)$. 
The $T_{cc}^+$ is found as a $D^0D^{*\,+}$ molecule ($87\%$) with a binding energy of $387$ keV/c$^2$ and a width of $81$ keV, in agreement with the experimental measurements.
The quark content of the state forces the inclusion of exchange diagrams to treat indistinguishable quarks between the $D$ mesons, which are found to be essential to bind the molecule.
The $D^0D^0\pi^+$ line shape, scattering lengths and effective ranges of the molecule are also analyzed, which are found to be in agreement with the LHCb analysis.
We search for further partners of the $T_{cc}^+$ in other charm and bottom sectors, finding different candidates. In particular, in the charm sector we find a shallow $J^P=1^+$ $D^+D^{*\,0}$ molecule ($83\%$), dubbed $T_{cc}^\prime$, just $1.8$ MeV above the $T_{cc}^+$ state. In the bottom sector, we find an isoscalar and an isovector $J^P=1^+$ bottom partners, as $BB^*$ molecules lying $21.9$ MeV/c$^2$ ($I=0$) and $10.5$ MeV/c$^2$ ($I=1$), respectively, below the $B^0B^{*\,+}$ threshold.
\end{abstract}


\keywords{Potential models, Quark models, Coupled-channels calculation}

\maketitle


\section{Introduction}
\label{sec:Introduction}

In the last two decades, the spectroscopy of heavy hadrons has showed an unexpected complexity compared to the predictions of naive quark models. The discovery of exotic tetraquark-like states such as the $X(3872)$~\cite{Belle:2003nnu}, $Z_c(3900)$~\cite{Ablikim:2013mio} and $Z_{cs}(3985)$~\cite{Ablikim:2020hsk}; or pentaquark-like structures like the $P_c(4450)^+$~\cite{LHCb:2015yax} and $P_{cs}(4459)^0$~\cite{LHCb:2020jpq}, has pushed forward our comprehension of Quantum Chromodynamics at this scale.

The LHCb Collaboration announced in 2021 the discovery of a new tetraquark-like state, named $T_{cc}^+$, with minimum quark content $cc\bar u\bar d$, close to the $D^{*+}D^0$ threshold in the $D^0D^0\pi^+$ invariant mass spectrum~\cite{LHCb:2021vvq}. Its importance relies on its doubly charmed content, in contrast with the hidden-charm nature of previous exotic states. Its measured mass and width, obtained from a Breit-Wigner parametrisation, with respect to the aforementioned $D^{*+}D^0$ threshold is
\begin{align}\label{eq:BWexp}
\delta m_{\rm BW} &= (-273\pm61\pm5^{+11}_{-14})\, {\rm keV/c}^2 \,\nonumber,\\
\Gamma_{\rm BW}   &= (410\pm165\pm43^{+18}_{-38})\,{\rm keV} \,.
\end{align}
where $\delta m_{\rm BW}=m_{T_{cc}^+}-m_{D^{*+}D^0}$.

In a further analysis performed by the LHCb collaboration about the structure of the $T_{cc}^+$ signal~\cite{LHCb:2021auc}, an estimation of the pole position in the complex energy plane provided the following values:  $\delta m_{\rm pole}=(-360\pm40^{+4}_{-0})$~keV/c$^2$ and $\Gamma_{\rm pole}=(48\pm2^{+0}_{-14})$~keV. That is to say, the mass is compatible with the previous measurement of Eq.~\eqref{eq:BWexp} but the width is much lower, which is reasonable due to the limited phase space of the  $D^0D^0\pi^+$ decay. This may indicate that the $\sim\!\!400$ keV width of Ref.~\cite{LHCb:2021vvq} is a consequence of the resolution function, whose root mean square is actually around $400$ keV~\cite{LHCb:2021vvq, LHCb:2021auc}. Thus, we have an extremely narrow state, very close to threshold, which is a strong candidate for a pure molecular state.

The $T_{cc}^+$ recalls the $X(3872)$, which is a candidate for a loosely-bound $D\bar D^*$+h.c. molecule; however, we are now dealing with an open-charmed state that radically changes its nature, making it explicitly exotic. In fact, the $T_{cc}^+$ state was already predicted $40$ years ago (see, \emph{e.g.}, Refs.~\cite{Ader:1981db, Zouzou:1986qh, Lipkin:1986dw, Heller:1986bt, Manohar:1992nd}). Its discovery has also inspired a sizable amount of theoretical studies~\cite{Meng:2022ozq, Maiani:2022qze, Lin:2022wmj, Cheng:2022qcm, Achasov:2022onn, Padmanath:2022cvl, Agaev:2022ast, Yang:2019itm, Deng:2021gnb, Santowsky:2021bhy, Du:2021zzh, Baru:2021ldu, Albaladejo:2021vln, Ren:2021dsi, Xin:2021wcr, Guo:2021yws, Ling:2021bir, Agaev:2021vur,Qin:2020zlg} that mostly associate such state to a $DD^*$ molecule or a compact tetraquark (see the review of Ref.~\cite{Chen:2022asf}, and references therein, for further details). Nevertheless, the proximity of the $DD^*$ threshold will mean that, although there may exist a compact tetraquark part, the dominant component will be the molecular one, so the description of this state as a pure $DD^*$ molecule is more than justified.

We explore herein the molecular nature of the $T_{cc}^+$ state as a $DD^*$ system using a constituent quark model (CQM)~\cite{Vijande:2004he, Segovia:2013wma} which has been widely used in heavy quark sectors, studying their spectra~\cite{Segovia:2009zz, Segovia:2015dia, Yang:2019lsg, Ortega:2020uvc}, their electromagnetic, weak and strong decays and reactions~\cite{Segovia:2011zza, Segovia:2011dg, Segovia:2014mca, Martin-Gonzalez:2022qwd}, and their coupling with meson-meson thresholds~\cite{Ortega:2016pgg, Ortega:2018cnm, Ortega:2021xst, Ortega:2021fem}, see also the reviews~\cite{Ortega:2020tng, Ortega:2012rs}. The advantage of using a model with relatively large history is that it allows us to make predictions. Indeed, all the parameters of the model have already been constrained from previous works. Then, from this perspective, we present a parameter-free calculation, and we are able to analyze in a rather robust way the properties of $T_{cc}^+$ and its possible partners.

The manuscript is arranged as follows. After this introduction, the theoretical framework is presented in Sec.~\ref{sec:Theory}. Section~\ref{sec:Results} is mostly devoted to the analysis and discussion of our theoretical results. Finally, we summarize and give some conclusions in Sec.~\ref{sec:Summary}.


\section{Theoretical formalism}
\label{sec:Theory}

The reader is referred to, for instance, Refs.~\cite{Segovia:2008zz, Segovia:2008zza, Segovia:2012cd, Ortega:2009hj} for the explicit expressions of all the potentials and model parameters involved in this calculation. We shall briefly focus here on the most relevant features of the mentioned quark model, and its generalization to describe meson-meson systems.

The rationale of the CQM lies on the emergence of a dynamical quark mass as a consequence of the spontaneous breaking of the chiral symmetry in QCD. From the Goldstone theorem, this symmetry breaking induces the appearance of boson exchanges between light quarks, phenomenology that can be modeled with the effective Lagrangian~\cite{Diakonov:2002fq}
\begin{equation}
{\mathcal L} = \bar{\psi}(i\, {\slash\!\!\! \partial}
-M(q^{2})U^{\gamma_{5}})\,\psi  \,,
\end{equation}
invariant under chiral transformations. This equation shows the dynamical constituent quark mass, $M(q^2)$, and the matrix of Goldstone-boson fields, $U^{\gamma_5} = e^{i\lambda _{a}\phi ^{a}\gamma _{5}/f_{\pi}}$. An expansion of such matrix gives
\begin{equation}
U^{\gamma _{5}} = 1 + \frac{i}{f_{\pi}} \gamma^{5} \lambda^{a} \pi^{a} -
\frac{1}{2f_{\pi}^{2}} \pi^{a} \pi^{a} + \ldots
\end{equation}
The first term represents the constituent light quark mass, the second encodes the one-boson exchange interaction between quarks and the last one illustrates a two-pion exchange term, which can be modeled as a scalar meson exchange potential. Goldstone-boson exchanges are missing if one of the interacting quarks is a heavy quark, that is, $QQ$ and $Qq$ pairs, being $Q=\{c,b\}$ and $q=\{u,d,s\}$, due to the explicit breaking of the chiral symmetry in such sectors.

Apart from this nonperturbative feature, the model adds a one-gluon exchange potential, given by the vertex Lagrangian
\begin{equation}
{\mathcal L}_{qqg} = i\sqrt{4\pi\alpha_{s}} \, \bar{\psi} \gamma_{\mu}
G^{\mu}_{c} \lambda^{c} \psi \,,
\label{Lqqg}
\end{equation}
where $\lambda^{c}$ are the $SU(3)$ color matrices, $G^{\mu}_{c}$ the
gluon field and $\alpha_{s}$ an effective strong coupling constant whose scale dependence (see its explicit expression in, \emph{e.g.}, Ref.~\cite{Segovia:2008zz}), allows a consistent description of light, strange
and heavy mesons.

Last, but not least, the model is completed with confinement in order to prevent from having colorful hadrons. The confining potential is a linearly-rising one with respect to the distance between quarks; however, at relatively large distances, it is screened due to the presence of sea quarks that induces the breaking of the $q\bar q$ binding string described by Ref.~\cite{Bali:2005fu},
\begin{equation}
V_{\rm CON}(\vec{r}\,)=\left[-a_{c}(1-e^{-\mu_{c}r})+\Delta \right]
(\vec{\lambda}_{q}^{c}\cdot\vec{\lambda}_{\bar{q}}^{c}) \,,
\label{eq:conf}
\end{equation}
where $a_{c}$ and $\mu_{c}$ are model parameters. The effective confinement strength of the linear behavior at short distances is $\sigma=-a_{c}\,\mu_{c}\,(\vec{\lambda}_{q}^{c}\cdot\vec{\lambda}_{\bar{q}}^{c})$, while it becomes constant at large distances with a plateau given by $\left[\Delta-a_{c} \right] (\vec{\lambda}_{q}^{c} \cdot \vec{\lambda}_{\bar{q}}^{c})$, above whose energy no $q\bar q$ bound states can be found.

From this underlying quark--(anti-)quark interaction, the $D$ meson spectrum is built. The masses and wave functions of the $D$ mesons are obtained from the Schr\"odinger equation, by means of the Gaussian Expansion Method~\cite{Hiyama:2003cu}. The GEM simplifies the evaluation of matrix elements while maintaining enough accuracy (see, e.g., Ref.~\cite{Ortega:2012rs} for details).

Within our theoretical formalism, the meson-meson interaction, needed for a molecular description of the $T_{cc}^+$, emerges from the microscopic description at the quark level using the Resonating Group Method (RGM)~\cite{Wheeler:1937zza}. That is, the cluster-cluster interaction is a residue of the interactions between quarks, which are described by the CQM potentials. 

\begin{figure}[!t]
\includegraphics[width=.5\textwidth]{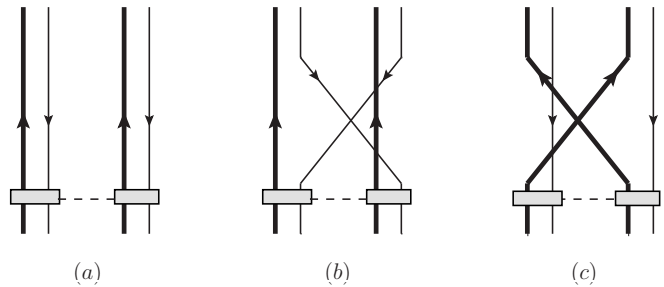}
\caption{\label{fig:diagrams} Scheme of diagrams that are included in this work. \emph{Panel (a)} Direct diagrams. \emph{Panel (b)} Exchange diagrams between light antiquarks. \emph{Panel (c)} Exchange diagrams between charm quarks. In all panels, the gray band represents the sum of interactions between quarks of different clusters (see Eqs.~\eqref{eq:directV} and~\eqref{eq:exchangeV} and related text); and the position of the charm quark is shown as a thick line.}
\end{figure}

For the $D^{(*)}D^{(*)}$ system under evaluation, the meson-meson interaction can be split in a direct kernel, with no quark rearrangement between clusters, and an exchange kernel, where such rearrangement are considered (see Fig.~\ref{fig:diagrams} for a schematic representation). Then, for a general transition $AB\to A'B'$, we have
\begin{align}\label{eq:fullpot}
 ^{\rm RGM}V(\vec P',\vec P) = \,^{\rm RGM}V_D(\vec P',\vec P) + \,^{\rm RGM}K(\vec P',\vec P) \,,
\end{align}
being $\vec P^{(\prime)}$ the initial (final) relative momentum of the $AB$ ($A'B'$) system, $^{\rm RGM}V_D$ the direct potential and $^{\rm RGM}K$ the exchange kernel, whose explicit expressions are shown below.

On the one hand, general direct potentials can be calculated as,
\begin{align}\label{eq:directV}
&
{}^{\rm RGM}V_{D}(\vec{P}',\vec{P}) = \sum_{i\in A, j\in B} \int d\vec{p}_{A'} d\vec{p}_{B'} d\vec{p}_{A} d\vec{p}_{B} \times \nonumber \\
&
\times \phi_{A'}^{\ast}(\vec{p}_{A'}) \phi_{B'}^{\ast}(\vec{p}_{B'})
V_{ij}(\vec{P}',\vec{P}) \phi_A(\vec{p}_{A}) \phi_B(\vec{p}_{B})  \,,
\end{align}
with $V_{ij}$ the interaction at quark level given by the CQM,  $\vec p_{A(B)}$ the relative internal momentum of the $A$ ($B$) meson and $i$ $(j)$ the indices that run inside the constituents of the $A$ ($B$) meson. We see that the wave functions of the mesons act as natural cut offs for the potentials, providing, at the same time, a domain of applicability of the model around the included channel thresholds of the same order.

On the other hand, the $D^{(*)}D^{(*)}$ system has indistinguishable quarks ($c\bar  q- c\bar q'$), so a full antisymmetric wave function must be built as

\begin{align}
 \Psi = {\cal A}\left[\phi_{A}\phi_{B}\chi_L\sigma_{ST}\xi_c\right]
\end{align}
where ${\cal A}$ is the full antisymmetric operator, $\phi_{A(B)}$ is the wave functions of the $A(B)$ meson, $\chi_L$ the relative orbital wave function of the $AB$ pair, $\sigma_{ST}$ their spin-isospin wave function and $\xi_c$ their color wave function.
Then, the $D^{(*)}D^{(*)}$ full antisymmetric operator is given, up to a normalization factor, as ${\cal A}=(1-P_q)(1-P_c)$, where $P_q$ is the operator that exchanges light antiquarks (Fig.~\ref{fig:diagrams}(b)) and $P_c$ the operator that exchanges charm quarks between clusters (Fig.~\ref{fig:diagrams}(c)).

The exchange kernel $^{\rm RGM}K$, that models the quark rearrangement between mesons, can be written as
\begin{align}
{}^{\rm RGM}K(\vec P',\vec P_i) &= {}^{\rm RGM}H_E(\vec P',\vec P_i) \nonumber \\
&
- E_T\,{}^{\rm RGM} N_E(\vec P',\vec P_i)\,.
\end{align}
That is, a non-local and energy-dependent kernel which can be separated in a potential term plus a normalization term. Here, $E_T$ is the total energy of the system and $\vec P_i$ is a continuous parameter. The exchange Hamiltonian and normalization can be written as
\begin{subequations}
\begin{align}\label{eq:exchangeV}
&
{}^{\rm RGM}H_{E}(\vec{P}',\vec{P}_{i}) = \int d\vec{p}_{A'}
d\vec{p}_{B'} d\vec{p}_{A} d\vec{p}_{B} d\vec{P} \phi_{A'}^{\ast}(\vec{p}_{A'}) \times \nonumber \\
&
\times  \phi_{B'}^{\ast}(\vec{p}_{B'})
{\cal H}(\vec{P}',\vec{P}) P_{12} \left[\phi_A(\vec{p}_{A}) \phi_B(\vec{p}_{B}) \delta^{(3)}(\vec{P}-\vec{P}_{i}) \right] \,,\\
&
{}^{\rm RGM}N_{E}(\vec{P}',\vec{P}_{i}) = \int d\vec{p}_{A'}
d\vec{p}_{B'} d\vec{p}_{A} d\vec{p}_{B} d\vec{P} \phi_{A'}^{\ast}(\vec{p}_{A'}) \times \nonumber \\
&
\times  \phi_{B'}^{\ast}(\vec{p}_{B'})
P_{12} \left[\phi_A(\vec{p}_{A}) \phi_B(\vec{p}_{B}) \delta^{(3)}(\vec{P}-\vec{P}_{i}) \right] \,,
\end{align}
\end{subequations}
where ${\cal H}$ is the Hamiltonian at quark level.

The mass and properties of the $T_{cc}^+$ molecular candidate obtained from the coupled-channels calculation can be extracted from the $T$ matrix, obtained from the Lippmann-Schwinger equation
\begin{align} \label{ec:Tonshell}
T_\beta^{\beta'}(z;p',p) &= V_\beta^{\beta'}(p',p)+\sum_{\beta''}\int dq\,q^2\,
V_{\beta''}^{\beta'}(p',q) \nonumber \\
&
\times \frac{1}{z-E_{\beta''}(q)}T_{\beta}^{\beta''}(z;q,p) \,,
\end{align}
where $\beta$ represents the set of quantum numbers necessary to determine a partial wave in the meson-meson channel, $V_{\beta}^{\beta'}(p',p)$ is the full RGM potential from Eq.~\eqref{eq:fullpot} and  $E_{\beta''}(q)$ is the energy for the momentum $q$ referred to the lower threshold.


\section{Results}
\label{sec:Results}

\subsection{$T_{cc}^+$ as a $DD^*$ molecule}

With the aim of evaluating the molecular nature of the $T_{cc}^+$, we perform a coupled-channels calculation of the $J^P=1^+$ $cc\bar q\bar q'$ sector, including the following meson-meson thresholds~\footnote{The threshold masses (in MeV/c$^2$) are shown in parenthesis.}: $D^0D^{*\,+}$ (3875.10), $D^+D^{*\,0}$  (3876.51) and $D^{*\,0}D^{*\,+}$ (4017.11) , which in the mentioned sector both mesons can be in relative $^3S_1$ or $^3D_1$ partial waves. As already mentioned before, we have a molecular system with a quark content of the kind $Q\bar  q- Q\bar q'$. The system recalls the $X(3872)$ problem, studied in Ref.~\cite{Ortega:2009hj} using the same formalism, but in this case there is explicit charm content. Therefore, we cannot couple the molecular system to charmonium states when studying the $T_{cc}^+$ structure; this would provide additional attraction in order to help binding the molecule as in the case of the X(3872). Nevertheless, besides the direct interaction between $D^{(*)}D^{(*)}$ pairs, we have to consider exchange diagrams to deal with indistinguishable quarks from different mesons in the molecule.

Indeed, the right mixture of mesons must be considered in order to have a well-defined isospin state. The quark content of $D$ mesons is $c\bar q$. Using the isospin doublets $D=(D^+,-D^0)$ and $D^*=(D^{*\,+},-D^{*\,0})$ the isospin basis is built as
\begin{align*}
| D^*D,I=0\rangle &= -\frac{1}{\sqrt{2}}\left(|D^{*\,+}D^0\rangle-|D^{*\,0}D^+\rangle\right) \,, \\
| D^*D,I=1\rangle &= -\frac{1}{\sqrt{2}}\left(|D^{*\,+}D^0\rangle+|D^{*\,0}D^+\rangle\right) \,.
\end{align*}
The energy difference between $D^0D^{*\,+}$ and $D^+D^{*\,0}$ is roughly 1.4~MeV, larger than the binding energy of the $T_{cc}^+$, so to explore possible isospin breaking effects of this threshold energy difference we will perform a calculation in charged basis, defined as
\begin{align*}
|D^{*\,0}D^+\rangle &= -\frac{1}{\sqrt{2}}\left(| D^*D,I=1\rangle-| D^*D,I=0\rangle\right) \,, \\
|D^{*\,+}D^0\rangle &= -\frac{1}{\sqrt{2}}\left(| D^*D,I=1\rangle+| D^*D,I=0\rangle\right) \,.
\end{align*}

The calculation shows one bound state below the lower $D^0D^{*\,+}$ threshold, with a mass with respect to the latter threshold of $-387$ keV/c$^2$, fully compatible with the experimental value of Ref.~\cite{LHCb:2021auc}, $(-360\pm 40^{+4}_{-0})$ keV/c$^2$, and less than $2\sigma$ from the one reported in Ref.~\cite{LHCb:2021vvq}, $(-273\pm61\pm5^{+11}_{-14})$ keV/c$^2$. Most of the attraction is due to pseudo-Goldstone boson exchanges, but the state is unbound unless the exchange kernel is considered. From this perspective, the rearrangement contribution is essential here, as it was the coupling to charmonium states in the case of the $X(3872)$~\cite{Ortega:2009hj}.

The probabilities of each channel are shown in Table~\ref{tab:props}. The state is basically a $D^0D^{*\,+}$ molecule, with $\sim\!\! 87\%$ probability due to its proximity to this threshold; the other $13\%$ corresponds to the $D^+D^{*\,0}$ channel. These probabilities are similar to the results of the coupled-channels analysis presented in Ref.~\cite{Albaladejo:2021vln}. Our molecule is essentially an $I=0$ state ($\sim\!\!81\%$), but the experimental mass difference between the $D^0D^{*\,+}$ and $D^+D^{*\,0}$ thresholds adds a sizable isospin breaking to the bound state, around $19\%$ of $I=1$. Note that, in our framework, the isospin breaking is only due to this mass difference because the potentials do not break isospin.

\begin{table}[!t]
\caption{\label{tab:props} Probabilities of each channel (in \%) for the bound states found in this work within the $J^P=1^+$ $cc\bar q\bar q'$ sector.}
\begin{ruledtabular}
\begin{tabular}{c|ccc|cc}
State & ${\cal P}_{D^0D^{*+}}$ & ${\cal P}_{D^+D^{*0}}$& ${\cal P}_{D^{*+}D^{*\,0}}$ & ${\cal P}_{I=0}$& ${\cal P}_{I=1}$   \\ \hline
$T_{cc}$ & $86.8$ & $13.1$ & $0.1$ & $81.3$ & $18.7$  \\
$T_{cc}^\prime$ & $16.9$ & $83.1$ & $0.01$ & $57.7$ & $42.3$  \\
\end{tabular}
\end{ruledtabular}
\end{table}

\begin{table}[!t]
\caption{\label{tab:widths} Properties of the bound states found in this work within the $J^P=1^+$ $cc\bar q\bar q'$ sector. The binding energy to the nearest threshold ($E_B=m_{\rm thres}-m_{T_{cc}}$) is given in keV/c$^2$, the pole position ($M-i\,\frac{\Gamma}{2}$) in MeV, and the partial widths are given in keV.  The partial widths includes the contributions from both the $D^0D^{*\,+}$ and $D^+D^{*\,0}$ channels, decaying through the unstable $D^{*\,+}$ and $D^{*\,0}$ meson, respectively.
}
\begin{ruledtabular}
\begin{tabular}{c|cc|ccc}
State &  $E_B$ & $M-i\,\frac{\Gamma}{2}$ & $\Gamma_{D^0D^0\pi^+}$ &$\Gamma_{D^0D^+\pi^0}$ &$\Gamma_{D^0D^+\gamma}$ \\
\hline
$T_{cc}$ & $387$& $3874.713$ &  $49$ & $26$ & $6$ \\
$T_{cc}^\prime$ & $3$ & $3876.507-i\,0.129$ & $175$ & $140$ & $40$ \\

\end{tabular}
\end{ruledtabular}
\end{table}

The pole position and partial widths are given in Table~\ref{tab:widths}. The  $T_{cc}^+$ state can only decay strongly if the meson $D^*$ inside the $DD^*$ system disintegrates. As the $D^*$ width is small, $\Gamma_{D^{*\,+}}=83.4\pm1.8$ keV~\cite{PDG2022}, the decay can be calculated perturbatively considering the $D^*$ an unstable particle decaying into $D\pi$ or $D\gamma$. That is to say, e.g., the decay $T_{cc}^+\to D^0D^0\pi^+$ can be calculated as
\begin{align}\label{eq:decay}
\Gamma_{D^0D^0\pi^+} &= \Gamma_{D^{*+}\to D^0\pi^+}\int_0^{k_{\rm max}} k^2dk |\chi_{D^0D^{*+}}(k)|^2 \nonumber \\ 
&
\times\frac{ (M_T-E_{D^0}-E_{D^{*+}})^2 }{(M_T-E_{D^0}-E_{D^{*+}})^2+\frac{\Gamma_{D^*}^2}{4}}\,,
\end{align}
where $\Gamma_{D^{*+}\to D^0\pi^+}$ is the $D^{*\,+}$ experimental partial width to $D^0\pi^+$, $\chi_{D^0D^{*+}}(k)$ is the wave function of the channel $D^0D^{*\,+}$, $E_D$ are the total energies of the mesons involved in the reaction and $k_{\rm max}$ is the maximum on-shell momentum of the $D^0D^0\pi^+$ system:
\begin{equation}
k_{\rm max}=\frac{1}{2M_T}\sqrt{\left[M_T^2-(2m_{D^0}+m_{\pi^+})^2\right]\left[M_T^2-m_{\pi^+}^2\right]} \,,
\end{equation}
where $M_T$ is the mass of the $T_{cc}^+$. The $D^0D^0\pi^+$ threshold is located at about $3869$ MeV, \emph{i.e.} there is not much phase space available, which explains the small partial width obtained.

Considering both the decays of the $D^{*\,+}$ and $D^{*\,0}$ mesons inside the $D^0D^{*\,+}$ and $D^+D^{*\,0}$ channels, respectively, and taking the $\Gamma_{D^{*\,0}}=\Gamma_{D^{*\,+}}=83.4$ keV, we obtain a total width of 81~keV (see Table~\ref{tab:widths}). This value is larger than the estimated one in Ref.~\cite{LHCb:2021auc}, $\Gamma_{\rm pole}=48\pm2^{+0}_{-14}$ keV. These values are compatible with other studies, \emph{e.g.}, the $78$ keV width of Ref.~\cite{Lin:2022wmj}, $61$ keV of Ref.~\cite{Cheng:2022qcm}, $63$ keV of Ref.~\cite{Ling:2021bir}, the range $43$ keV to $80$ keV of Ref.~\cite{Feijoo:2021ppq} and $\sim78$ keV of Ref.~\cite{Albaladejo:2021vln}. However Ref.~\cite{Baru:2021ldu} showed that the width of the state is sensitive to three body effects, so we have performed a similar calculation introducing the energy dependent self-energy of the $D^*$ meson. In this calculation the binding energy goes to 278 keV while the width drops to 42.1 keV, in better agreement with LHCb analysis. Notice that in contrast to Ref.~\cite{Baru:2021ldu} we don't have a counter term to adjust the real part of the binding energy and for this reason the effect on the real part looks more important in our case. Also, this is not a completely consistent calculation since we have not included the form factor introduced by the internal wave functions of the mesons in the calculation of the self energy. However, it shows that the disagreement of the
width is due to three body effects as stated in Ref.~\cite{Baru:2021ldu}.

Besides the $T_{cc}^+$ below $D^0D^{*\,+}$, we also find a molecular candidate for the $T_{cc}^\prime(3876)$ predicted in Ref.~\cite{Chen:2021vhg} slightly below the $D^+D^{*\,0}$ threshold in the $J^P=1^+$ $cc\bar q\bar q'$ sector. Its pole position is $3876.507-i\,0.129$ MeV, thus the binding energy with respect to the $D^+D^{*\,0}$ threshold is just $3$ keV/c$^2$. It has a width of $355$ keV,  summing up the partial widths from $D^0D^0\pi^+$ (175 keV), $D^0D^+\pi^0$ (140 keV) and $D^0D^+\gamma$ (40 keV). The properties of the bound state are detailed in the second line of Table~\ref{tab:props} and Table~\ref{tab:widths}.

There are two kinds of theoretical uncertainties in our results: one is inherently connected to the numerical algorithm and the other is related to the way of fixing the model parameters.  The numerical error is negligible. As mentioned before, the set of model parameters are fitted to reproduce a certain number of hadron observables within a determinate range of agreement with experiment. Therefore, it is difficult to assign an error to those parameters and, as a consequence, to the magnitudes calculated when using them. As the range of agreement between theory and experiment is around 10-20\% when dealing with physical observables that help to fix the model parameters (such as binding energies, mass splittings or decays), this range of agreement can be taken as an estimation of the model uncertainty for the derived quantities.

It is true that, for the $T_{cc}$ and specially the $T_{cc}^\prime$, having so small binding energies one would expect to be more sensitive to the longer range of the interaction, given
in our case by the one-pion-exchange interaction, which is the most model independent piece of our quark potential model. 
However, we cannot give a definitive answer on the existence of the $T'_{cc}$, although the conclusion
would be to have a state very close to threshold giving an small signal in the line shape.

\begin{figure}[!t]
\includegraphics[width=.5\textwidth]{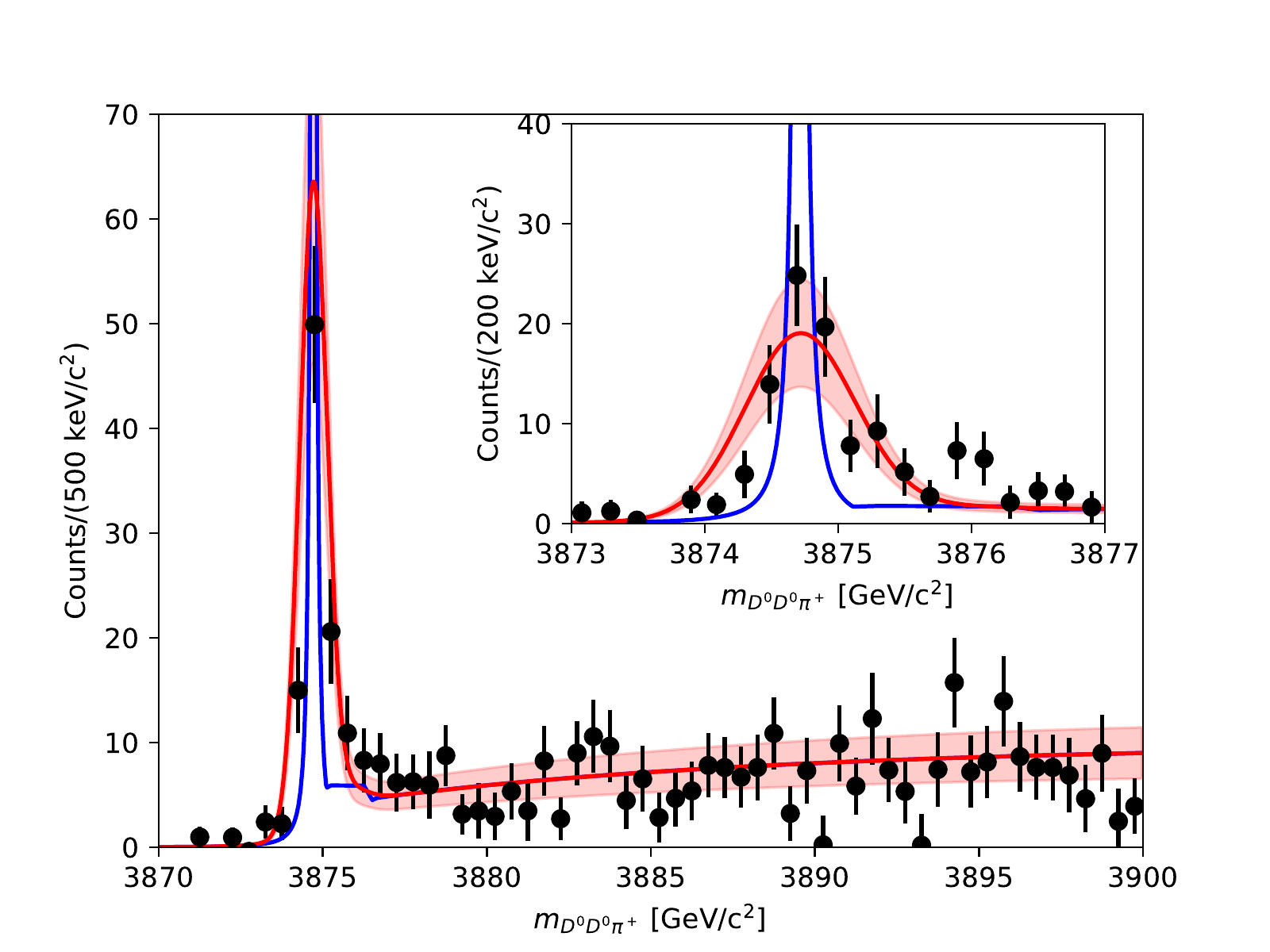}
\caption{\label{fig:line} Theoretical line shape of the $T_{cc}^+$ peak. The black solid point with error bars are the experimental data from Ref.~\cite{LHCb:2021auc}. The blue line is the theoretical line shape when the pole parameters of Table~\ref{tab:widths} are considered. The red line shows the theoretical line shape convoluted with the experimental resolution function, whereas the shadowed area shows the normalization uncertainty from the $\chi^2$ minimization.}
\end{figure}

The theoretical position of the pole and the width allows us to describe the experimental line shape with good accuracy (see Fig.~\ref{fig:line}). Although the predicted width is much smaller than the experimental data in Ref.~\cite{LHCb:2021auc}, when the resolution function is considered with a root mean square of $400$ keV~\cite{LHCb:2021vvq, LHCb:2021auc}, the predicted line shape agrees with the data. The $T_{cc}$ peak is clearly visible, while the $T_{cc}^\prime$ peak appears as a small bump smeared by the resolution, 
due to its closeness to the $D^+D^{*\,0}$ threshold, that enhanced this component against the $D^0D^{*+}$. The normalization of the curve is the only free parameter, which has been calculated with a $\chi^2$-minimization procedure.

\subsection{Scattering lengths and effective ranges}

In Table~\ref{tab:scat} we show the scattering length and effective range of each channel, taking the relation between the T matrix and phaseshifts $\delta_n$ for channel $n$ as

\begin{align}\label{eq:ERES}
 e^{2\,i\,\delta_n} = 1-\,i\,2\pi\mu_nk_n T_{nn}
\end{align}
being $k_n$ the on-shell momentum  and $\mu_n$ the reduced mass for channel $n$. 
And from there the usual effective range expansion,
\begin{align}\label{eq:ERE}
 k_n\,{\rm cotan}(\delta_n) = \frac{1}{a_n} + \frac{1}{2}r_nk_n^2 \,.
\end{align}

From Eq.~\eqref{eq:ERES} and~\eqref{eq:ERE}, at $k\to 0$ the scattering length has the expression

\begin{equation}\label{eq:defscatlen}
 a_n = -\pi\mu_n T_{nn}(E_{\rm th})
\end{equation}
with $T_{nn}$ evaluated at threshold (e.g., $E_{\rm th}=m_D+m_{D^*}$ for channel $DD^*$) for stable mesons. The effect of the width of the $D^*$ in the $T_{cc}^+$ can be studied by using a complex $D^*$ mass $m_{D^*}\to m_{D^*}-i\,\Gamma_{D^*}/2$ in Eq.~\eqref{ec:Tonshell}. Then, the on-shell momentum is replaced by~\cite{Baru:2021ldu,Gao:2018jhk,Du:2021zzh}

\begin{equation}
 k=\sqrt{2\mu (E-M_{\rm th}+i\Gamma/2)}
\end{equation}
so the $T$-matrix is evaluated at $E_{\rm th}=m_D+m_{D^*}-i\Gamma_{D^*}/2$ in Eq.~\eqref{eq:defscatlen}.
With this definition of Eq.~\eqref{eq:defscatlen} for the unstable $D^*$, the scattering length is almost real and independent on the value of the width of $D^{*\,0}$ and $D^{*\,+}$, as can be seen in Table~\ref{tab:scatLength}. The effect of the width, nevertheless, is evident if we compare $-\pi\mu_n T_{nn}(m_D+m_{D^*})$ values, which are evaluated at real energies different from the complex two-body threshold branch point.

The scattering length of the lower threshold $D^0D^{*\,+}$ is $-7.14$ fm, fully compatible with the experimental estimation of Ref.~\cite{LHCb:2021auc}, $a_{\rm sc}^{\rm LHCb}=-7.15(51)$ fm.  This value is larger than the scattering lengths calculated by Lattice-QCD in, e.g., Refs.~\cite{Chen:2022vpo, Ikeda:2013vwa,Padmanath:2022cvl}.  However, these lattice computations are made with large pion masses, due to the complexities of working with near physical pion mass, so a direct comparison with our scattering length result is meaningless at this moment. Future lattice calculations could improve these values. For example, Ref.~\cite{Padmanath:2022cvl} finds the $T_{cc}^+$ as a virtual state, but in the limit of physical light quark mass it is expected to become a real bound state pole.

Turning now our attention to the effective range, the LHCb collaboration only gives an upper limit of $r_{\rm exp}>-11.9(16.9)$ fm at 90(95)\% CL. Our calculation throws a value of $-0.49$ fm, again compatible with the experimental lower limit.

It is worth noticing that for $D^+D^{*\,0}$ and $D^{*\,0}D^{*\,+}$ the scattering length and effective range are complex numbers. The imaginary part of the scattering lengths and effective ranges indicate the existence of inelastic channels, which in this calculation are the lower thresholds, and can be related to the total inelastic scattering cross section~\cite{BALAKRISHNAN19975}. The $D^0D^{*\,+}$ scattering length is real because no lower threshold is considered in this work, though the effect of the width is small (see Table~\ref{tab:scatLength}). Alternatively, a larger imaginary part would be present if evaluated at the real threshold, as in the LHCb analysis of Ref.~\cite{LHCb:2021auc}. The real part of the scattering length of the $D^+D^{*\,0}$ is large and negative, compatible with the bound state $T_{cc}^\prime(3876)$, which has a small binding energy as detailed in previous section. Alternatively, the scattering length of the $D^{*\,0}D^{*\,+}$ channel is small and positive, while its effective range has a large imaginary part, which should be taken with caution.

\begin{table}[!t]
\caption{\label{tab:scat} Scattering lengths, effective ranges and couplings of the channels considered in this work. See text for details. }
\begin{ruledtabular}
\begin{tabular}{cccc}
Channel  & $a_{\rm sc}$ [fm] & $r_{\rm eff}$ [fm] & $g$ [GeV$^{-1/2}$] \\
\hline
$D^0D^{*\,+}$  & $-7.14$ & $-0.49$ & $0.12$ \\
$D^+D^{*\,0}$ & $-8.98+8.57\,i$ & $0.82 + 0.48\,i$ & $0.07$ \\
$D^{*\,0}D^{*\,+}$ & $0.20+0.02\,i$  & $-6.09-6.23\,i$ & $<0.01$ \\
\end{tabular}
\end{ruledtabular}
\end{table}

\begin{table*}[!t]
\caption{\label{tab:scatLength} $D^0D^{*\,+}$ pole
position and S-wave scattering lengths for coupled-channels calculation with an unstable $D^*$ meson, considering different $D^{*\,0}$ and $D^{*\,+}$ widths. We distinguish between the scattering length evaluated at the real
$M_{D^0D^{*\,+}}=m_{D^0}+m_{D^{*\,+}}$ (third column) or at the complex
$E_{D^0D^{*\,+}}=m_{D^0}+m_{D^{*\,+}}-i\Gamma_{D^{*\,+}}/2$ (forth
column). See discussion after Eq.~\eqref{eq:ERE}.}
\begin{ruledtabular}
\begin{tabular}{l|ccc}
Case & $E_R-i\,\Gamma_R/2$ &$-\pi\mu T(M_{D^0D^{*\,+}})$ & $a_{\rm
sc,D^0D^{*\,+}}$ \\ \hline
$\Gamma_{D^{*\,0}}=\Gamma_{D^{*\,+}}=0$  & $3874.713-i\,0$ &
$-7.14+0.00\,i$         & $-7.14+0.00\,i$\\
$\Gamma_{D^{*\,0}}=0$, $\Gamma_{D^{*\,+}}=83.4$ keV &
$3874.713-i\,0.036$& $-8.64+2.32\,i$ & $-7.14-0.08\,i$ \\
$\Gamma_{D^{*\,0}}=\Gamma_{D^{*\,+}}=83.4$ keV & $3874.713-i\,0.042$ & 
$-8.58+2.43\,i$ & $-7.14+0.001\,i$ \\
\hline
\end{tabular}
\end{ruledtabular}
\end{table*}

Finally, close to the $T_{cc}^+$, the $T$ matrix can be written as
\begin{align}
 T_{ij}\sim \frac{g_ig_j}{E-M_T}
\end{align}
where $M_T$ is the mass of the $T_{cc}^+$ state and $g_i$ is the coupling of the state to the $i$-th channel.
In Table~\ref{tab:scat} we show the coupling constants to the $D^0D^{*\,+}$, $D^+D^{*\,0}$ and $D^{*\,0}D^{*\,+}$ channels. We see that the coupling to $D^0D^{*\,+}$ and $D^+D^{*\,0}$ channels are of the same order, which indicates that the isospin breaking shown in Table~\ref{tab:props} is basically due to the mass difference of the thresholds, as the interactions conserve isospin.


\subsection{$T_{cc}^+$ partners and the bottom sector}

We have searched for $T_{cc}^+$ partners in alternative spin-parity sectors and thresholds,~\footnote{Note that our prediction of partners is made within the same theoretical approach without fine-tuning and thus associated model uncertainties should apply here, even more cautiously since we add thresholds located far away from the $T_{cc}$ pole mass.} such as the $DD$ system in $J^P=0^+$ or $D^*D^*$ with quantum numbers $J^P=0^+$, $1^+$ and $2^+$. We have not found additional bound states. In particular, we have searched for pure isospin-1 partners in the $D^+D^{*\,+}$ channel in $J^P=1^+$, but we did not find any state, or peak signal in the $D^+D^0\pi^+$ channel. We do, however, find a virtual and a resonance in $J^P=0^+$ in isospin $1$ and $0$, respectively, just below the $D^*D^*$ threshold, in a coupled-channels calculation of the $DD+D^*D^*$ channels.  Additionally, in $J^P=1^+$, besides the $T_{cc}$ and $T_{cc}^\prime$, we find a pole in the second Riemann sheet just below the $D^*D^*$ threshold. Their properties can be found in Table~\ref{tab:moreTcc}.


\begin{table}[!t]
\caption{\label{tab:moreTcc} Properties of the $T_{cc}$ candidates as $D^*D^*$ molecules in the $J^P=0^+$ and $1^+$ sectors obtained in this work. Masses, widths, binding energies (defined as $E_B=M_T-M_{D^*D^*}$) and partial widths are shown in MeV/c$^2$.}
\begin{ruledtabular}
\begin{tabular}{cc|ccccc}
$J^P$ & $I$ & Mass & Width & $E_B$ &  ${\cal P}_{D^*D^*}$ & Type\\
\hline
\multirow{2}{*}{$0^+$} & $0$     & $4018.0$ & $8.15$ & $0.9$ & $95.6\%$ & Resonance \\\cline{2-7}
 & $1$    & $4016.9$ & $0.6$ & $-0.2$  & $98.8\%$ & Virtual \\ \hline
 $1^+$ & $-$ & $4014.0$ & $0$ & $-3.1$ & $38.5\%$ & Virtual \\
\end{tabular}
\end{ruledtabular}
\end{table}

\begin{table}[!t]
\caption{\label{tab:propsTb} Mass and binding energy (in MeV/c$^2$) and probabilities of each channel (in \%) for the $J^P=1^+$ $T_{bb}$ states predicted in this work.}
\begin{ruledtabular}
\begin{tabular}{cc|ccc|cc}
Mass & $E_B$ & ${\cal P}_{B^0B^{*+}}$ & ${\cal P}_{B^+B^{*0}}$& ${\cal P}_{B^{*+}B^{*\,0}}$ & ${\cal P}_{I=0}$& ${\cal P}_{I=1}$\\
\hline
$10582.2$ & $21.9$ & $47.8$ & $50.0$ & $2.2$ & $99.99$ & $0.01$ \\
$10593.5$ & $10.5$ & $51.0$ & $48.6$ & $0.4$ & $0.02$ & $99.98$ \\
\end{tabular}
\end{ruledtabular}
\end{table}

We have, then, expanded our analysis to the bottom sector, that is, states with $\bar b\bar b u d$ minimum quark content. First, we analyzed the $J^P=1^+$ sector in a coupled-channels calculation analog to that of the $T_{cc}^+$, that is, in charged basis including the $B^0B^{*+}$, $B^+B^{*0}$ and $B^{*+}B^{*\,0}$ thresholds. We find two bound states below the $B^0B^{*+}$ channel, an isoscalar and an isovector state. The isoscalar $T_{bb}$ state has a mass of $10582.2$ MeV/c$^2$ and a binding energy of $21.9$ MeV/c$^2$, whereas the isovector $T_{bb}$ state has a mass of $10593.5$ MeV/c$^2$ and a binding energy of $10.5$ MeV/c$^2$. The large binding energies represent a challenge to detect these state, but decays through the $B^*\to B\gamma$ should be detectable, as the $T_{bb}$ molecule lies above the $BB\gamma$ threshold. Besides, the isospin breaking is negligible in both cases and these states are pure $I=0$ and $I=1$ molecules. The probabilities of each channel for both $J^P=^+$ states are shown in Table~\ref{tab:propsTb}, being balanced between the $B^0B^{*\,+}$ and $B^+B^{*\,0}$ channels.

\begin{figure}[!t]
\includegraphics[width=.45\textwidth]{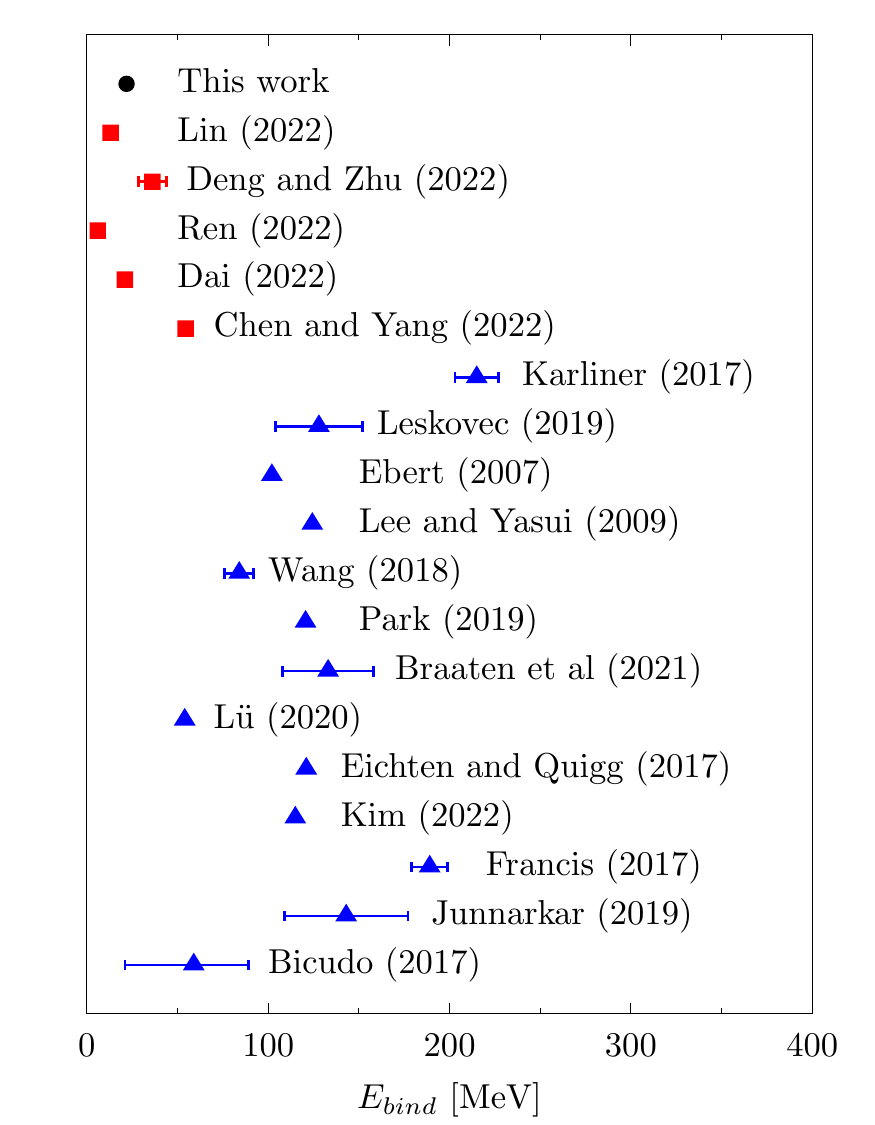}
\caption{\label{tab:Tbbworks} Binding energy $E_{\rm bind}=M_{BB^*}-M_{T_{bb}}$ of the $(I)J^P=(0)1^+$ $T_{bb}$ predicted by this work in comparison with a selection of previous studies, using the threshold value $M_{BB^*}=10604$ MeV/c$^2$. Black circle: this work; Red squares: molecular candidates from Refs.~\cite{Lin:2022wmj,Deng:2021gnb,Ren:2021dsi,Dai:2022ulk,Chen:2021tnn}; Blue triangles: tetraquark candidates from Refs.~\cite{Karliner:2017qjm,Leskovec:2019ioa,Ebert:2007rn,Lee:2009rt,Wang:2017uld,Park:2018wjk,Braaten:2020nwp,Lu:2020rog,Eichten:2017ffp,Kim:2022mpa,Francis:2016hui,Junnarkar:2018twb,Bicudo:2016ooe}. }
\end{figure}

In Fig.~\ref{tab:Tbbworks}, we compare our isoscalar $T_{bb}$ molecule with the predictions from other theoretical studies. Our binding energy agrees with those reported by molecular descriptions, whereas the tetraquark studies usually show much deeper bound states.

\begin{table}[!t]
\caption{\label{tab:moreTbb} Properties of the $T_{bb}$ candidates as $B^{(*)}B^{(*)}$ molecules in the $J^P=0^+$ and $2^+$ sectors obtained in this work. Masses, widths, binding energies and partial widths are shown in MeV/c$^2$.}
\begin{ruledtabular}
\begin{tabular}{cc|ccccccc}
$J^P$ & $I$ & Mass & Width & $E_B$ & ${\cal P}_{BB}$ & ${\cal P}_{B^*B^*}$ & $\Gamma_{BB}$& $\Gamma_{B^*B^*}$\\
\hline
\multirow{4}{*}{$0^+$} & \multirow{2}{*}{$0$}     & $10553.0$ & 0 & $6.0$ & $92\%$ & $8\%$ & 0 & 0 \\
 &   & $10640.7$ & $2.8$ & $8.7$ & $76\%$ & $24\%$ & $2.8$ & 0 \\ \cline{2-9}
 & \multirow{2}{*}{$1$}    & $10545.9$ & 0 & $13.1$ & $93\%$ & $7\%$ & 0 & 0 \\
 &  &  $10672.6$ & $72.0$ & $-23.2$ & $39\%$ & $61\%$ & $30.7$ & $41.3$\\ \hline
 $2^+$ & $1$ & $10642.3$ & 0 & $7.1$ & - & $100\%$ & - & 0 \\
\end{tabular}
\end{ruledtabular}
\end{table}

Besides the $BB^*$ $J^P=1^+$ molecules, we have searched for further $T_{bb}$ states in the $J^P=0^+$ and  $2^+$ sectors, including all the meson-meson channels in a relative $S$ wave, that is, $BB+B^*B^*$ for $0^+$ and $B^*B^*$ for $2^+$.
In this case, we have performed separate calculations for $I=0$ and $I=1$, as the isospin breaking was found to be negligible for the $1^+$ $T_{bb}$.
We find the five candidates shown in Table~\ref{tab:moreTbb}. These results show a populated spectroscopy in the bottom sector, which can be detected in future searches.

\section{Summary}
\label{sec:Summary}

Within the framework of a constituent quark model, widely used to describe other exotic structures in the heavy meson and baryon spectrum, we have explored the molecular assignment of the recently-discovered $T_{cc}^+$ state, an open-charmed system with $cc\bar u\bar d$ minimum quark content. The calculation describes the $DD^*$ and $D^*D^*$ systems by means of pseudo-Goldstone bosons exchanges, one-gluon interactions and a linear-screened confinement potential between quarks, including both direct and quark exchange diagrams as shown in Fig.~\ref{fig:diagrams}.

We performed a coupled-channels calculation of the $J^P=1^+$ $cc\bar q\bar q'$ sector in charged basis, considering the $D^0D^{*\,+}$, $D^+D^{*\,0}$ and $D^{*\,0}D^{*\,+}$ channels, and find a molecular candidate with a mass with respect to the lower $D^0D^{*\,+}$ threshold of $-387$ keV/c$^2$ and a width of $81$ keV, in agreement with the experimental measurements of the $T_{cc}^+$. This structure is only bound when the quark exchange diagrams are included. The molecule is a $\sim\!\!87\%$ $D^0D^{*\,+}$ system with a sizable isospin breaking ($\sim\!\!81\%$ of $I=0$), only due to the mass difference of the $D^0D^{*\,+}$ and $D^+D^{*\,0}$. With these parameters, the experimental line shape is perfectly described. We have also analyzed the scattering length and effective ranges of the molecule, which are in agreement with the LHCb analysis of Ref.~\cite{LHCb:2021auc}.

Regarding the possibility of having $T_{cc}^+$ partners, we find a candidate for the $T_{cc}^\prime(3876)$ slightly below the $D^+D^{*\,0}$ threshold in $J^P=1^+$, just $1.8$ MeV above the $T_{cc}^+$ state. We do not find any other molecular structure in the charm sector. We do, however, find several bottom partners in the $J^P=0^+$, $1^+$ and $2^+$ $\bar b\bar b ud$ sectors. In the $J^P=1^+$ sector we find an isoscalar and isovector $BB^*$ molecule. The isoscalar $T_{bb}^+$, the heavy partner of the $T_{cc}^+$, is predicted to lie $21.9$ MeV/c$^2$ below the $B^0B^{*\,+}$ threshold, which agrees with other molecular predictions in the literature.


\begin{acknowledgments}
This work has been partially funded by
EU Horizon 2020 research and innovation program, STRONG-2020 project, under grant agreement no. 824093;
Ministerio Espa\~nol de Ciencia e Innovaci\'on, grant no. PID2019-107844GB-C22 and PID2019-105439GB-C22;
and Junta de Andaluc\'ia, contract nos. P18-FR-5057 and Operativo FEDER Andaluc\'ia 2014-2020 UHU-1264517.
\end{acknowledgments}

\appendix

\section{Relation of OPE in the $X(3872)$ and $T_{cc}^+$ channels.}

There is some controversy between the relation between the one-pion-exchange interaction
in the channels of the $X(3872)$ and the $T_{cc}^+$. In Refs.~\cite{Tornqvist1994,PhysRevD.78.034007} the authors predict opposite overall signs for the $DD^*$ and $D\bar D^*$ systems,
but the same relative sign for the OPE is obtained for positive C-parity. However, in those references
there is no explicit mention to the antisymmetrization of the wave function for the $DD^*$,~\footnote{In other words, a $(PV+VP)/\sqrt{2}$ combination is taken in those references, with $P$ for pseudoscalar and $V$ for vector mesons.} which changes
the sign for different isospin. In order to clarify this controversy, we derive the relation in this appendix.

The $X(3872)$ is a $D\bar D^*$ plus $D^*\bar D$ state with $J^{PC}=1^{++}$ quantum numbers.
We use for mesons (and antimesons) $q\bar q$ wave functions~\footnote{Note that PDG uses for
antimesons $\bar q q$ states.} so we write the wave function of a meson as
\begin{align}
|r_{12} JM;LSf_1\bar f_2 \rangle =&
\sum_{M_L M_S} (LM_LSM_S|JM) R(r_{12}) 
\nonumber \\ &
Y_{LM_L}(\hat r_{12}) 
|s_1 s_2; S M_S\rangle |f_1 \bar f_2 \rangle
\end{align}
with $J$, $M$, $L$, $S$ the total angular momentum, three component of the total angular momentum,
orbital angular momentum and spin quantum numbers of the meson, respectively, and $r_{12}$ is the relative
coordinate of the two quarks. Notice that we omit the color degree of freedom since it does not play any role.

The $C$-parity operator transforms
the state $|f_1 \bar f_2\rangle$ in a state $|f_2 \bar f_1\rangle$ with the same quantum numbers
and a phase given by $(-1)^{L+S}$ so with our convention
\begin{align}
	C|D\rangle &= |\bar D \rangle
	\nonumber \\
	C|D^*\rangle &= -|\bar D^* \rangle
\end{align}
The two-meson $AB$ wave function is, then, written as
\begin{align}
|AB;J_TM_TJ_{AB} L \rangle
=&
\sum_{M_{AB},M}
(LMJ_{AB}M_{AB}|J_TM_T) 
\nonumber \\ & 
\sum_{M_A,M_B}
(J_AM_AJ_BM_B|J_{AB}M_{AB})
\nonumber \\ & 
R_{AB}(r_{1234})
Y_{LM}(\hat r_{1234})
\nonumber \\ & 
|r_{12} J_{A}M_{A};L_{A}S_{A}f_1\bar f_2 \rangle 
\otimes \nonumber \\ & \otimes
|r_{34} J_{B}M_{B};L_{B}S_{B}f_3\bar f_4 \rangle
\end{align}
coupled to $J_T$, $M_T$, $J_{AB}$ and $L$ total quantum numbers and $r_{1234}$ is the
relative coordinate of the $12$ quark pair with the $34$ pair. With this convention the $C$ parity
operator gives
\begin{align}
C|AB;J_TM_TJ_{AB} L \rangle =&
(-1)^{J_{AB}-J_A-J_B+L}
\nonumber \\ &
|C(B)C(A);J_TM_TJ_{AB} L \rangle
\quad 
\end{align}
This means that the $J^P=1^+$ states with well-defined $C$-parity are
\begin{align}
|\Psi_{D\bar D^*}^\pm \rangle =& \frac{1}{\sqrt 2} (
|D\bar D^* \rangle \mp
|D^*\bar D \rangle )
\nonumber \\
C|\Psi_{D\bar D^*}^\pm \rangle =& 
\pm |\Psi_{D\bar D^*}^\pm \rangle 
\nonumber
\end{align}
The central part of OPE potential between the light $q\bar q$ pair is
\begin{align}
V_{23}(q) =& \frac{1}{(2\pi)^3} \frac{g_{ch}^2}{4m_q^2} \frac{\Lambda^2}{\Lambda^2+q^2}
\frac{q^2}{q^2+m^2}
\frac 1 3 (\vec \sigma_2 \cdot \vec \sigma_3) 
(\vec \tau_2 \cdot \vec \tau_3)
\end{align}
for a $c\bar n-n\bar c$ ordering.

Since for $\bar q$ we use the $G$-parity convention, the isospin factor is just
$\langle \vec \tau_2 \cdot \vec \tau_3 \rangle = 2I(I+1)-3$. The potential
at meson level is obtained as
\begin{align}
&\langle \Psi_{D\bar D^*}^\pm | V_{23} |\Psi_{D\bar D^*}^\pm \rangle =
( 2I(I+1)-3)
\frac{1}{(2\pi)^3} \frac{g_{ch}^2}{4m_q^2} \nonumber \\ &\times\frac 1 3
\int d^3 p'_{12} d^3 p'_{34} d^3 p_{12} d^3 p_{34}
\delta^3(p'_1-p_1)
\delta^3(p'_4-p_4)
\nonumber \times \\ &\times
\delta^3(p'_2+p'_3-p_2-p_3)
\frac{\Lambda^2}{\Lambda^2+q_{23}^2} \frac{q_{23}^2}{q_{23}^2+m^2}
\nonumber \times \\ &\times
\big( R_D(p'_{12})R_{D^*}(p'_{34})R_D(p_{12})R_{D^*}(p_{34})
\langle 01 | \vec \sigma_2 \cdot \vec \sigma_3 | 01 \rangle \mp
\nonumber \\& \mp
R_D(p'_{12})R_{D^*}(p'_{34})R_{D^*}(p_{12})R_{D}(p_{34})
\langle 01 | \vec \sigma_2 \cdot \vec \sigma_3 | 10 \rangle \big)
\nonumber 
 \end{align}
where $\langle 01 | \vec \sigma_2 \cdot \vec \sigma_3 | 01 \rangle = 0$ and
$\langle 01 | \vec \sigma_2 \cdot \vec \sigma_3 | 10 \rangle =-1$.
Although the calculations are performed with the $D^{(*)}$ meson wave functions
solutions of the Schr\"odinger equation, in order to evaluate the sign we can derive it for
simple gaussians
\begin{align}
        R_D(p) = R_{D^*}(p) =& \left( \frac{1}{2\pi\beta} \right)^{3/4} e^{-\frac{p^2}{4\beta}}
\end{align}
and we obtain
\begin{align}
\langle \Psi_{D\bar D^*}^\pm | V_{23} |\Psi_{D\bar D^*}^\pm \rangle &=
\pm ( 2I(I+1)-3) V(Q)
e^{-\frac{Q^2}{16\beta}}\quad 
\nonumber \\
\end{align}
with $Q$ the transferred momentum  between the two mesons and
\begin{align}
V(q) =&
\frac{1}{(2\pi)^3} \frac{g_{ch}^2}{4m_q^2} \frac 1 3
\frac{\Lambda^2}{\Lambda^2+q^2} \frac{q^2}{q^2+m^2} 
\end{align}
which shows that the tail of OPE is repulsive for $I=0$ as seen in Fig.~1
of Ref.~\cite{PhysRevD.78.034007}.

For the channel of the $T_{cc}^+$ the situation is different, since now quarks are
identical and we have to include antisymmetry between them. 
We have that, for the quark ordering $c\bar n-c\bar n$, the antisymmetrizer is given by\footnote{Here we write $P_c=P_{13}$
and $P_q=P_{24}$ to have explicitly which are the quarks (antiquarks) that are exchanged.}
\begin{align}
        {\mathcal A} = (1-P_{24})(1-P_{13})
\end{align}
Introducing the operator $P=P_{13}P_{24}$ that exchange mesons it can be written as
\begin{align}
        {\mathcal A} = (1-P_{24})(1-P)
\end{align}
So the wave functions is
\begin{align}\label{eq:antisym}
|{\mathcal A}(AB);J_TM_TJ_{AB} L \rangle &=
{\mathcal A}|AB;J_TM_TJ_{AB} L \rangle
\nonumber \\ &= (1-P_{24}) \big[
|AB;J_TM_TJ_{AB} L \rangle
\nonumber \\ & 
+(-1)^\mu |BA;J_TM_TJ_{AB} L \rangle \big]
\nonumber \\
\end{align}
with $\mu=L+S-S_A-S_B+I-I_A-I_B$. 

For the $T_{cc}^+$ channel, $J^P=1^+$,  $\mu=I-1$, hence, the global sign is $(-1)^\mu=-1$ for $I=0$ and $(-1)^\mu=1$ for $I=1$. This means that, for $I=0$, we have to take the
states
\begin{align}
|\Psi_{DD^*} \rangle &= \frac{1}{\sqrt 2} (
|DD^* \rangle -
|D^*D \rangle )
\end{align}
Now the interaction is between the two light antiquarks so
\begin{align}
V_{24}(q) &= -\frac{1}{(2\pi)^3} \frac{g_{ch}^2}{4m_q^2} \frac{\Lambda^2}{\Lambda^2+q^2}
\frac{q^2}{q^2+m^2}
\frac 1 3 (\vec \sigma_2 \cdot \vec \sigma_4) 
(\vec \tau_2 \cdot \vec \tau_4)
\end{align}
and again the isospin factor is just
$\langle \vec \tau_2 \cdot \vec \tau_4 \rangle = 2I(I+1)-3$. The potential for the direct term (we exclude $P_{24}$ in Eq.~\eqref{eq:antisym}
which gives the exchange kernels) is
\begin{align}
&\langle \Psi_{DD^*} | V_{24} |\Psi_{DD^*} \rangle =
-( 2I(I+1)-3)
\frac{1}{(2\pi)^3} \frac{g_{ch}^2}{4m_q^2} \nonumber \\ &\times
\frac 1 3
\int d^3 p'_{12} d^3 p'_{34} d^3 p_{12} d^3 p_{34}
\delta^3(p'_1-p_1)
\delta^3(p'_3-p_3)
\nonumber \times \\ &\times
\delta^3(p'_2+p'_4-p_2-p_4)
\frac{\Lambda^2}{\Lambda^2+q_{24}^2} \frac{q_{24}^2}{q_{24}^2+m^2}
\nonumber \times \\ &\times
\big( R_D(p'_{12})R_{D^*}(p'_{34})R_D(p_{12})R_{D^*}(p_{34})
\langle 01 | \vec \sigma_2 \cdot \vec \sigma_4 | 01 \rangle +
\nonumber \\& +
(-1)^\mu R_D(p'_{12})R_{D^*}(p'_{34})R_{D^*}(p_{12})R_{D}(p_{34})
\langle 01 | \vec \sigma_2 \cdot \vec \sigma_4 | 10 \rangle \big)
\nonumber 
 \end{align}
where $\langle 01 | \vec \sigma_2 \cdot \vec \sigma_4 | 01 \rangle = 0$ and
$\langle 01 | \vec \sigma_2 \cdot \vec \sigma_4 | 10 \rangle =1$.
Performing the same calculation we get
\begin{align}
\langle \Psi_{DD^*} | V_{24} |\Psi_{DD^*} \rangle &=
(-1)^{\mu+1}( 2I(I+1)-3) V(Q)
e^{-\frac{Q^2}{16\beta}}\quad 
\end{align}
with $\mu=I-1$ for $J^P=1^+$ $DD^*$, which shows that the $(I)J^P=(0)1^+$ $DD^*$ channel has the same sign as $(I)J^P=(0)1^+$ $D\bar D^*$ channel with $C=+$, and opposite sign 
than that with $C=-$.


\bibliography{Tcc-letter}

\end{document}